\documentclass[letterpaper,twocolumn,10pt]{article}
\usepackage{usenix2019_v3}

\usepackage{tikz}
\usepackage[normalem]{ulem}
\usepackage{amsmath,amssymb,amssymb,mathtools}
\usepackage{multirow,booktabs}
\usepackage[linesnumbered,ruled,vlined]{algorithm2e}
\usepackage{float}
\usepackage{graphicx}
\usepackage{subfig}
\usepackage{mathrsfs}
\usepackage{filecontents}
\usepackage[available,functional,reproduced]{usenixbadges}
\newcommand{\sysname}{{Llumnix}}
\newcommand{\syslet}{{llumlet}}
\newcommand{\eg}{\textit{e.g.}}
\newcommand{\ie}{\textit{i.e.}}
\newcommand{\etc}{\textit{etc}}
\newcommand{\aka}{\textit{a.k.a.}}
\renewcommand{\paragraph}[1]{{\bf \noindent #1 \hspace{5pt}}}

\newcommand{\revise}[1] {\textcolor{black}{#1}}
\newcommand{\final}[1] {\textcolor{black}{#1}}

\newenvironment{packed_itemize}
{
  \begin{list}{\labelitemi}{\leftmargin=1.0em}
    \setlength{\itemsep}{4pt}
    \setlength{\parskip}{0pt}
    \setlength{\parsep}{0pt}
    \setlength{\headsep}{0pt}
    \setlength{\topskip}{0pt}
    \setlength{\topsep}{0pt}
    \setlength{\partopsep}{0pt}
  }{\end{list}}

\begin{document}

\date{}

\title{\Large \bf {\sysname}: Dynamic Scheduling for Large Language Model Serving}


\author{
{\rm Biao Sun\thanks{Equal contribution.}\hspace*{2mm}\thanks{Work done during internship at Alibaba Group.}\hspace*{2mm}, Ziming Huang\footnotemark[1]\hspace*{2mm}\footnotemark[2]\hspace*{2mm}, Hanyu Zhao\footnotemark[1]\hspace*{2mm}, Wencong Xiao, Xinyi Zhang\footnotemark[2]\hspace*{2mm}, Yong Li, Wei Lin}\\ \\
Alibaba Group
}

\maketitle


\begin{abstract}
Inference serving for large language models (LLMs) is the key to unleashing their potential in people's daily lives.
However, efficient LLM serving remains challenging today because the requests are inherently heterogeneous and unpredictable in terms of resource and latency requirements, as a result of the diverse applications and the dynamic execution nature of LLMs. Existing systems are fundamentally limited in handling these characteristics and cause problems such as severe queuing delays, poor tail latencies, and SLO violations.

We introduce \sysname{}, an LLM serving system that reacts to such heterogeneous and unpredictable requests by \emph{runtime rescheduling} across multiple model instances.
Similar to context switching across CPU cores in modern operating systems, \sysname{} reschedules requests to improve load balancing and isolation, mitigate resource fragmentation, and differentiate request priorities and SLOs. \sysname{} implements the rescheduling with an efficient and scalable live migration mechanism for requests and their in-memory states, and exploits it in a dynamic scheduling policy that unifies the multiple rescheduling scenarios elegantly.
Our evaluations show that \sysname{} improves tail latencies by an order of magnitude, accelerates high-priority requests by up to 1.5$\times$, and delivers up to 36\% cost savings while achieving similar tail latencies, compared against state-of-the-art LLM serving systems. \sysname{} is publicly available at \url{https://github.com/AlibabaPAI/llumnix}.
\end{abstract}

\section{Introduction}

Large language models (LLMs) such as the GPT series~\cite{gpt3, openai2023gpt4} are bringing generative AI to an unprecedented level. Their human-level generation capabilities are being quickly adopted in a wide range of domains, inspiring many imaginations for future applications, and expected to have profound influences on how people live and work.

Inference serving of LLMs plays a key role in LLM-powered services, becoming a critical workload in datacenters. Such services are typically backed by multiple instances of the LLM deployed on a GPU cluster. The system involves a \emph{scheduler} and an \emph{inference engine}, where a request is first dispatched by the scheduler to a model serving instance, then gets executed by the inference engine inside.
The requests are typically batched for execution on each instance to increase throughput and cost efficiency.


We observe unique characteristics of LLMs that 
call for new design philosophy of the serving infrastructure.
The first is \emph{workload heterogeneity}.
LLMs are designed to be universal, by learning as much knowledge as possible from whatever domains. 
People can query the same LLM in totally different situations or even build custom applications atop LLMs for various scenarios; for all of these, a context-specific input (\ie, prompt) is all you need~\cite{gpt3}.
Such universality and application diversity lead to heterogeneity of the inference requests, in terms of input lengths, output lengths, expected latencies, \etc.
For instance, the task of summarizing long text can introduce significant input lengths, where the latency of returning the first token (word) is often important to user experience~\cite{ringattention2023}.

The second characteristic is \emph{execution unpredictability.}
Serving an LLM request needs to run the model for multiple iterations,
each producing a single output token;
however, it is not known a priori how many tokens will be generated eventually.
Moreover, the iterative generation also brings considerable GPU memory consumption that dynamically grows with the tokens.
As such, the execution time and the resource demand of a request are both unpredictable.

These characteristics make an LLM inherently a \emph{multi-tenant} and \emph{dynamic} environment, serving heterogeneous and unpredictable workloads on multiple instances.
\revise{This behavior is fundamentally different from traditional DNN models, where the requests are homogeneous and the execution is one-shot, stateless, and deterministic. Instead, we find LLMs more similar to} modern operating systems hosting processes with dynamic working sets and different priorities on multiple cores.
Managing such systems has complex goals, which goes beyond what existing inference serving systems are designed for.
\revise{Although there has been a series of LLM-tailored inference engines that shows superior performance, such systems concentrate on the sole goal of maximizing throughput within a single instance~\cite{faster2023,orca2022,vllm2023}.
The request scheduling across instances, on the other hand, has received relatively little attention; the common practice today is still to use generic scheduling systems or policies inherited from the era of traditional DNNs~\cite{deepspeedmii,romero2021infaas,gujarati2020serving,triton,alphaserve2023}.}
Such a clear gap introduces challenges in the following aspects that are crucial in multi-tenant environments and online services.

{\noindent \bf Isolation.}
The system can hardly provide performance isolation to requests as their memory consumption grows unpredictably. Memory contention incurs performance interference and even preemptions of certain requests in a batch~\cite{vllm2023}, leading to highly unstable latencies and service-level objective (SLO) violations, significantly sacrificing user experiences.

{\noindent \bf Fragmentation.}
The varying request lengths and memory demands inevitably result in memory fragmentation across instances, which introduces conflicting scheduling objectives.
\revise{The running requests prefer load balancing to reduce preemptions and interference, but such load balancing fragments the free memory space across instances at the same time. 
The fragmentation can cause long queuing delays of new requests that instead require a large space on one instance for the input sequences.}
This conflict is difficult for the scheduler to reconcile with unpredictable arrivals and lengths of requests.

{\noindent \bf Priorities.}
Requests from different applications and scenarios naturally come with different latency objectives.
Online chatbots~\cite{chatgpt,bard} are interactive applications and are therefore with tight SLO constraints.
On the contrary, offline applications, such as evaluation~\cite{instructgpt2022}, scoring~\cite{holistic2022}, or data wrangling~\cite{wrangle2022}, are less sensitive to latency. \revise{Such different latency objectives are also a consequence of the commercial purpose of earning more profits from LLMs via diversified service classes (\eg, ChatGPT Plus \cite{chatgptplus}).}
However, existing LLM inference systems~\cite{orca2022, vllm2023} often treat all requests for a model equally and cannot differentiate their priorities, which has limitations in meeting different latency objectives of requests.

\begin{figure}[t]
    \centering
    \includegraphics[width=0.95\columnwidth]{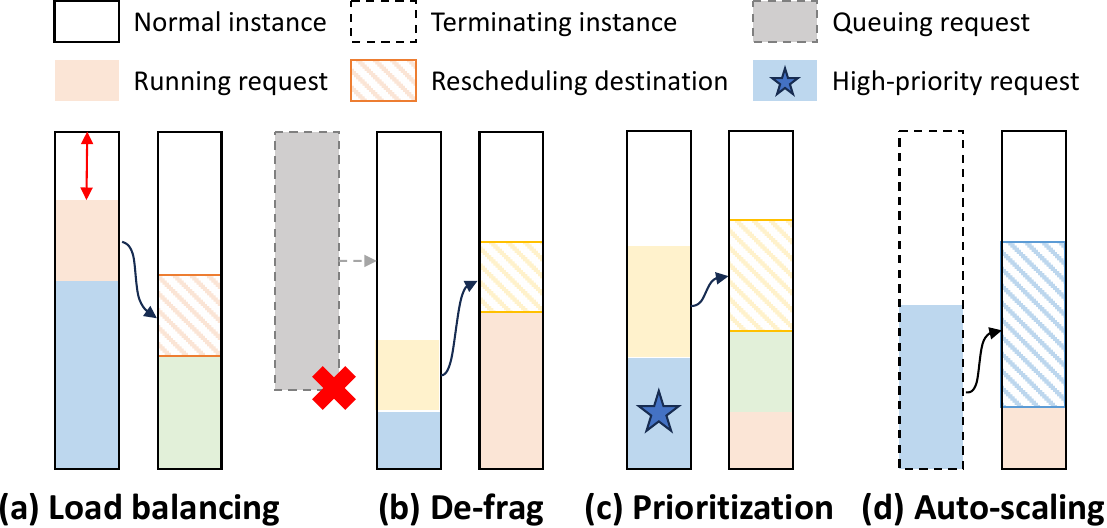}
    \caption{Example rescheduling scenarios in \sysname{}.}
    \label{fig:migration-cases}
\end{figure}

We introduce \sysname{}, a new \revise{scheduling system for LLM serving} that addresses the challenges above via \emph{runtime rescheduling} of requests across model instances.
Analogous to context switching across CPU cores in OS process management, \revise{rescheduling enables \sysname{} to \emph{react} to the unpredictable workload dynamics at runtime, instead of having to address all the complex scheduling concerns and tradeoffs with the one-shot dispatching of requests. \sysname{} reschedules requests for multiple purposes (Figure~\ref{fig:migration-cases}):} load balancing for reducing preemptions and interference, de-fragmentation for mitigating queuing delays, prioritization of urgent requests by creating even higher degree of isolation, saturating or draining out instances during auto-scaling more quickly.


\sysname{} reschedules requests via an efficient and scalable \emph{live migration} mechanism of requests along with their GPU memory states across instances. 
Straightforward rescheduling approaches could introduce substantial downtimes to rescheduled requests, especially for long sequences.
By contrast, \sysname{} introduces \emph{near-zero} downtime that is \emph{constant} to sequence lengths, by carefully coordinating the computation and the memory transfer to hide the cost.

To exploit such great scheduling flexibility of migration, \sysname{} adopts a distributed scheduling architecture that enables continuous rescheduling with high scalability. \sysname{} further introduces a dynamic scheduling policy under this architecture that unifies all the rescheduling scenarios with different goals elegantly. This unification is achieved via a concept called \emph{virtual usage}:
\sysname{} just needs to define a set of rules for setting the virtual usages of GPU memory for requests in different scenarios, and then use a simple load-balancing policy based on the virtual usages.

\revise{We have implemented \sysname{} as a scheduling layer on top of inference engines. \sysname{} currently supports a representative system, vLLM~\cite{vllm2023}, as the underlying engine.}
Evaluation on a 16-GPU cluster using realistic workloads shows that \sysname{} improves P99 first-token latency by up to 15$\times$ and P99 per-token generation latency by up to 2$\times$, compared against a state-of-the-art scheduler INFaaS~\cite{romero2021infaas}. \sysname{} also accelerates high-priority requests by 1.5$\times$, and achieves 36\% cost saving when delivering similar tail latencies.

In summary, this paper makes the following contributions.

\begin{packed_itemize}

\item We reveal the unique characteristics and scheduling challenges of LLM serving that necessitate new scheduling goals such as isolation, de-fragmentation, and priorities. 
\item We propose request rescheduling as a key measure to achieve these goals and realize it with an efficient migration mechanism of requests and their GPU memory states.
\item We design a distributed scheduling architecture and an accompanying scheduling policy that exploit request migration to achieve the multiple goals in a unified manner.
\item We implement and evaluate \sysname{} to show its advantages over state-of-the-art inference serving systems.

\end{packed_itemize}


\section{Background}

\paragraph{Application diversity of LLMs.}
Recent LLMs are becoming \emph{task-agnostic}. That is, the same model can work for various tasks with context-specific inputs (\aka~the ``prompts'') provided. This is achieved by both increasingly larger model and dataset sizes and advanced pre-training approaches such as few-shot learning~\cite{brown2020language}. Task-agnostic models enable diverse applications, from chatbots, search engines, summarization, coding, AI assistants, to AI agents, to name a few. 

The diverse applications lead to requests with different requirements for the serving. An important aspect is the \emph{sequence lengths}.
LLMs are racing to support longer sequence lengths ---
for example, from March to November 2023, the maximum sequence lengths of the GPT family have scaled from 32k \footnote{1k stands for 1,024 when describing sequence length in this paper.} (GPT-4~\cite{openai2023gpt4}) to 128k (GPT-4 Turbo~\cite{gpt4turbo2023}).
We expect this trend to continue as longer sequences are necessary for broader applications of LLMs. Consider an intuitive example of the tasks for summarizing and writing an article: they require sufficiently long input and output lengths, respectively.
Another aspect is \emph{expected latencies}. A real product example is that OpenAI introduces a subscription plan called ChatGPT Plus~\cite{chatgptplus} to offer faster responses of common ChatGPT services. In general, different applications and situations also naturally have different levels of urgency. For example, more interactive applications like personal assistants expect shorter latencies than tasks like summarizing an article.


\begin{figure}[t]
    \centering
    \includegraphics[width=0.85\columnwidth]{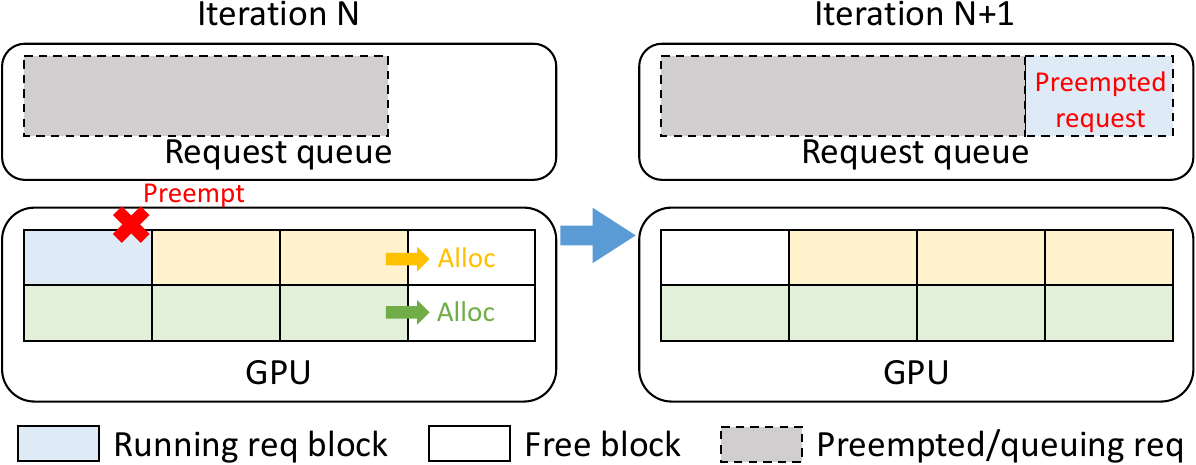}
    \caption{Request queuing and preemption using continuous batching and dynamic memory allocation.}
    \label{fig:pagedattn}
\end{figure}

\paragraph{Autoregressive generation.}
The inference for state-of-the-art LLMs is \emph{autoregressive}: the model iteratively accepts the input sequence plus all the previous output tokens to generate the next output token, until an ``end-of-sequence'' (EOS) token is generated. The phase for generating the first token and that for each new token afterwards are usually referred to as \emph{prefill} and \emph{decode}, respectively. LLM services typically return the generated tokens in a streaming manner. Therefore, the prefill and decode latencies are both user-perceivable and important to user experiences. The prefill latency determines how long it takes to start receiving the response, which can be dominated by the queuing delay. The decode latency determines the speed of receiving the following tokens subsequently.

During the autoregression, the intermediate results (key and value tensors used in the attention operation~\cite{vaswani2017attention}) for each token are involved in the generation of all following tokens. Therefore,
the inference engine typically stores these states in GPU memory for reuse, known as the \emph{KV cache}~\cite{pope2022efficiently}.

\paragraph{Batching and memory management.}
State-of-the-art inference engines apply the \emph{continuous batching} technique~\cite{orca2022,vllm2023} to handle the varying sequence lengths and dynamic arrivals of requests. That is, a new/completed request can join/leave the running batch immediately, instead of waiting for all the running requests to complete.
Batching also raises concern about memory management of KV cache.
Since the memory demand of KV cache is not known a priori, it would clearly limit the batch size and batching benefits if the memory is reserved to the maximum length.
For example, a LLaMA-2-13B~\cite{touvron2023llama2} model supports sequence lengths up to 4k, which translates to 3.2 GB KV cache for a single request; while the memory of current GPUs remain tens of GBs, let alone the space for model weights (26 GB for LLaMA-2-13B).
Therefore, recent work (vLLM~\cite{vllm2023}) proposed \emph{dynamic memory allocation} for KV cache to increase batch size and throughput, enabled by a technique named PagedAttention:
the KV cache tensors are stored in dynamically allocated blocks as the KV cache grows.
Figure \ref{fig:pagedattn} presents an example of using continuous batching with dynamic memory allocation. The running requests are chosen based on the free memory blocks, hence there is a queuing request (the gray one) at iteration N as the memory is insufficient. At the next iteration, the system runs out of memory for the new blocks of the running requests. Therefore, the system preempts certain running requests (the blue one), which then goes back to the queue.

\begin{figure}[t]
    \centering
    \includegraphics[width=0.95\columnwidth]{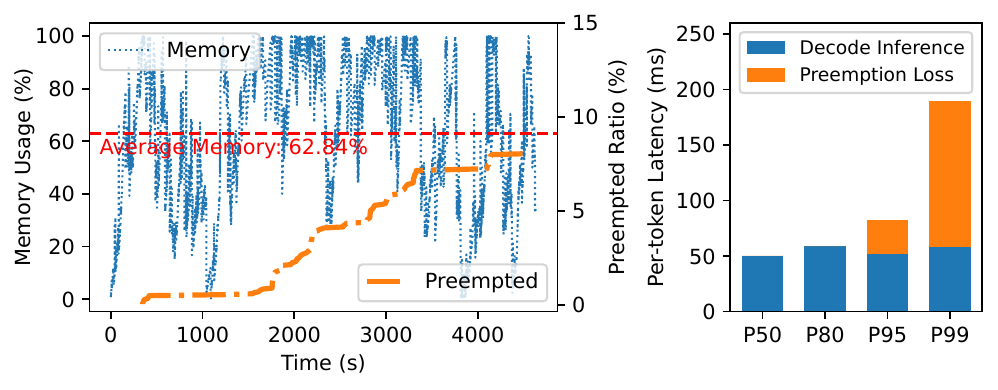}
    \caption{Request preemptions in LLaMA-7B serving.}
    \label{fig:preemption}
\end{figure}

\section{Motivation}



We motivate the design of \sysname{} with a series of key characteristics of LLM serving as follows.

\paragraph{Unpredictable memory demands and preemptions.}
With dynamic memory allocation, request preemptions are inevitable as a result of the unpredictable memory demands, which can significantly increase the latencies of the preempted requests.
Figure \ref{fig:preemption} shows an experiment of LLaMA-7B model serving using vLLM on an A10 GPU running a trace of 2,000 requests generated from a Poisson distribution. \revise{The input and output lengths follow a power-law distribution with a mean value of 256 tokens (details in \S\ref{sec:eval}).} We control the request rate (0.42 req/s) to get a moderate memory load (62\% on average) with some spikes due to the varying sequence lengths. Under such load, we still observe 8\% of the requests being preempted. We quantify the preemption loss by measuring the latency penalty caused by preemption, including the extra queuing time and the recomputing for previous KV cache. We show different percentiles of per-token decode latency (averaged across all decode iterations of a request).
We do not use the end-to-end latency because it depends on the number of iterations. We observe that the P99 per-token decode latency is much worse than the P50 (3.8$\times$), and the preemption loss accounts for 70\% for the P99 request. \revise{In particular, the P99 request experiences a total preemption loss of 50 seconds (preempted twice), showing severe service stalls and degradation of user experiences due to preemptions.}

\begin{figure}[t]
    \centering
    \includegraphics[width=0.7\columnwidth]{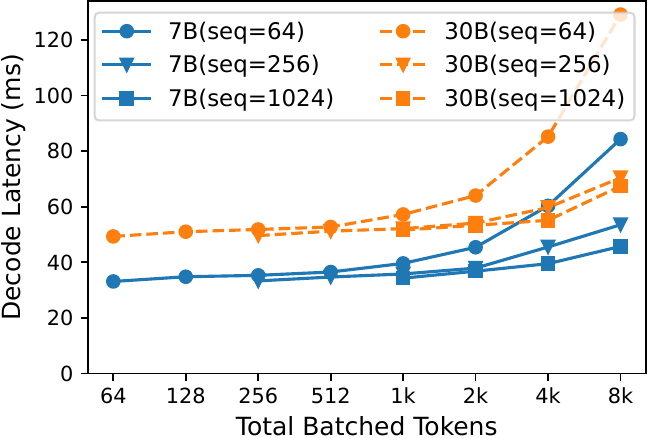}
    \caption{Latencies of one decode step of LLaMA-7B and LLaMA-30B with different sequence lengths and batch sizes.}
    \label{fig:interference}
\end{figure}

\paragraph{Performance interference among requests.}
We also observe performance interference of requests in a batch to each other, due to resource competition on GPU compute and memory bandwidth resources.
Figure \ref{fig:interference} shows the times for a decode step of LLaMA-7B (1-GPU) and LLaMA-30B (4-GPU) using different sequence lengths and batch sizes (the X-axis shows the total number of tokens in a batch for each data point). The decode speed decreases with more requests and higher interference, and the gap between the same sequence length is up to 2.6$\times$.

\paragraph{Memory fragmentation.}
Considering the aforementioned problems, it would be better to spread requests across instances to reduce preemptions and interference. However, such spreading will make the available memory of the cluster fragmented across instances simultaneously. 
\revise{
Here fragmentation refers to \emph{external} fragmentation, \ie, unallocated memory on an instance.
Dynamic allocation techniques like PagedAttention~\cite{vllm2023} can eliminate 
external fragmentation during the \emph{decode} phase, where the blocks are allocated one at a time. However, external fragmentation remains a significant problem for the \emph{prefill} phase, which requires many blocks on an instance in one allocation to accommodate the KV cache of all tokens in the inputs.
Therefore, external fragmentation can cause long queuing delays of new requests, especially those with long inputs.}

\begin{figure}[t]
    \centering
    \includegraphics[width=0.75\columnwidth]{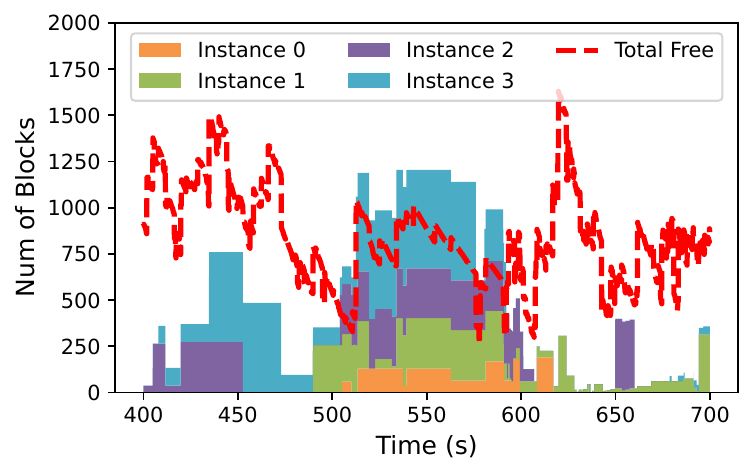}
    \caption{Total free memory vs. demands of the head-of-line queuing requests across four LLaMA-7B instances.}
    \label{fig:frag}
\end{figure}

Figure \ref{fig:frag} shows an experiment of four LLaMA-7B instances, \revise{where the trace also uses the input/output length distribution with mean value 256 and a Poisson distribution with a request rate of 1.9 req/s. 
We implement a spreading dispatching policy that dispatches new requests to the instance with the lowest memory load for load balancing.} We demonstrate the fragmentation by showing the total free memory blocks across the cluster, against the demand of the head-of-line queuing request on each instance. For most of the time span, the total free memory can accommodate the queuing requests on at least three instances (sometimes all of them). The request are queuing despite enough total memory because they exceed the free space on their own instances, which demonstrates the fragmentation and also the potential of de-fragmentation to reduce queuing delays.

\paragraph{Different emergency and priorities of requests.}
With requirements of products like ChatGPT Plus and the diverse application scenarios of LLMs, we foresee more applications with different latency sensitivities.
\revise{However, existing systems usually treat all requests equally, where the latency-sensitive could easily be interfered by other normal ones, \eg, excessive queuing delays or performance interference.} This calls for a systematic approach to differentiating the request priorities for an LLM to meet their respective latency objectives.




\paragraph{Opportunity: request rescheduling across instances.}
This paper explores a new dimension that is missing in current LLM serving systems: the multiple model instances of a deployment and their interaction. A simple intuition is that when the aforementioned problems occur on a certain instance, it is possible that the whole cluster still has enough space for avoiding preempting requests, accommodating new requests, or mitigating interference. \revise{This is also a natural consequence of the varying request lengths and memory loads across instances.} However, existing systems cannot exploit such free space on other instances because requests are tied on the same instance once scheduled \revise{throughout the autoregressive execution}. 
\sysname{} unifies the request scheduling component and the model inference engine
to explore the potentials of fine-grained coordination among inference instances.

\section{\sysname{} Design}


\subsection{Overview}

\sysname{} builds upon the key idea of rescheduling LLM inference requests at runtime across model instances.
\sysname{} inherits continuous batching~\cite{orca2022} and dynamic memory allocation~\cite{vllm2023} from state-of-the-art systems for high throughput.
Beyond that, \sysname{} exploits request rescheduling to react to the unpredictable workload dynamics in various situations with different scheduling goals, as illustrated in Figure \ref{fig:migration-cases}.

A first goal is \emph{load balancing} (Figure \ref{fig:migration-cases}-a) to reduce request preemptions and interference on high-load instances.
Although the dispatching can also consider load balancing of memory usage, it could be sub-optimal as the final memory usages of requests are unknown at the arrivals, due to the unpredictability of output lengths. 
Rescheduling complements it by reacting to the real usage growths of requests.
Meanwhile, as shown before, load balancing can also lead to higher memory fragmentation and longer queuing delays of long inputs probably.
Therefore, \sysname{} also reschedules requests for \emph{de-fragmentation} (\ref{fig:migration-cases}-b), \ie, creating contiguous space on an instance by moving requests onto others. Although these two goals remain a tradeoff, \sysname{} has a much larger space to balance them with rescheduling.
Another goal is \emph{prioritization} (\ref{fig:migration-cases}-c) of certain requests by rescheduling co-located requests away for lower load and avoiding interference. Such rescheduling provides ``decicated'' resources to high-priority requests dynamically, without the need for reserving machines statically.
Finally, \sysname{} also reschedules requests during \emph{auto-scaling}, \eg, to drain out an instance to be terminated (\ref{fig:migration-cases}-d) or saturate a new instance more quickly.

Realizing such highly dynamic rescheduling efficiently is challenging, considering the large request context states (\ie, the KV cache). 
Na\"ive solutions include recomputing or copying the KV cache of the rescheduled requests, however with high computation stalls and downtime, reaching over 50$\times$ of the decoding cost (\S\ref{eval:migrate}).
What's more, the KV cache states increase with sequence lengths, limiting the scheduling flexibility under the trend of growing context lengths~\cite{gpt4turbo2023}.
Such a high inference delay in generating next tokens greatly degrades the user experiences of LLM serving and thus prohibits request rescheduling.
\sysname{} addresses this challenge with a \emph{live migration} mechanism that pipelines and coordinates the KV cache copying and the token generation computation, thereby bringing negligible downtime (\S\ref{sec:migration}).


To exploit the benefits of migration, \sysname{} adopts a scalable architecture that combines global and local scheduling to decentralize the scheduling decisions and the coordinated migration actions, facilitating continuous rescheduling at scale (\S\ref{sec:arch}).
Under this architecture, we further design an efficient heuristic scheduling policy that centers around the \emph{virtual usage} concept to abstract the requirements of the different scheduling goals in a unified manner (\S\ref{sec:policy}).

\subsection{Live Migration of LLM Requests}
\label{sec:migration}
The significant KV cache states of requests can potentially introduce great cost and serving stalls during rescheduling. 
\sysname{} addresses this challenge by exploiting a key characteristic of LLM inference: the KV cache is \textit{append-only}. LLM inference iteratively concatenates the output token of the current iteration with the input tokens, which is set as the input for the next iteration. In this way, inference engines also keep appending the calculated KV state of the current iteration to the KV cache parameters, leaving the parameters generated by previous iterations remain constant.

\begin{figure}[t]
    \centering
    \includegraphics[width=0.99\columnwidth]{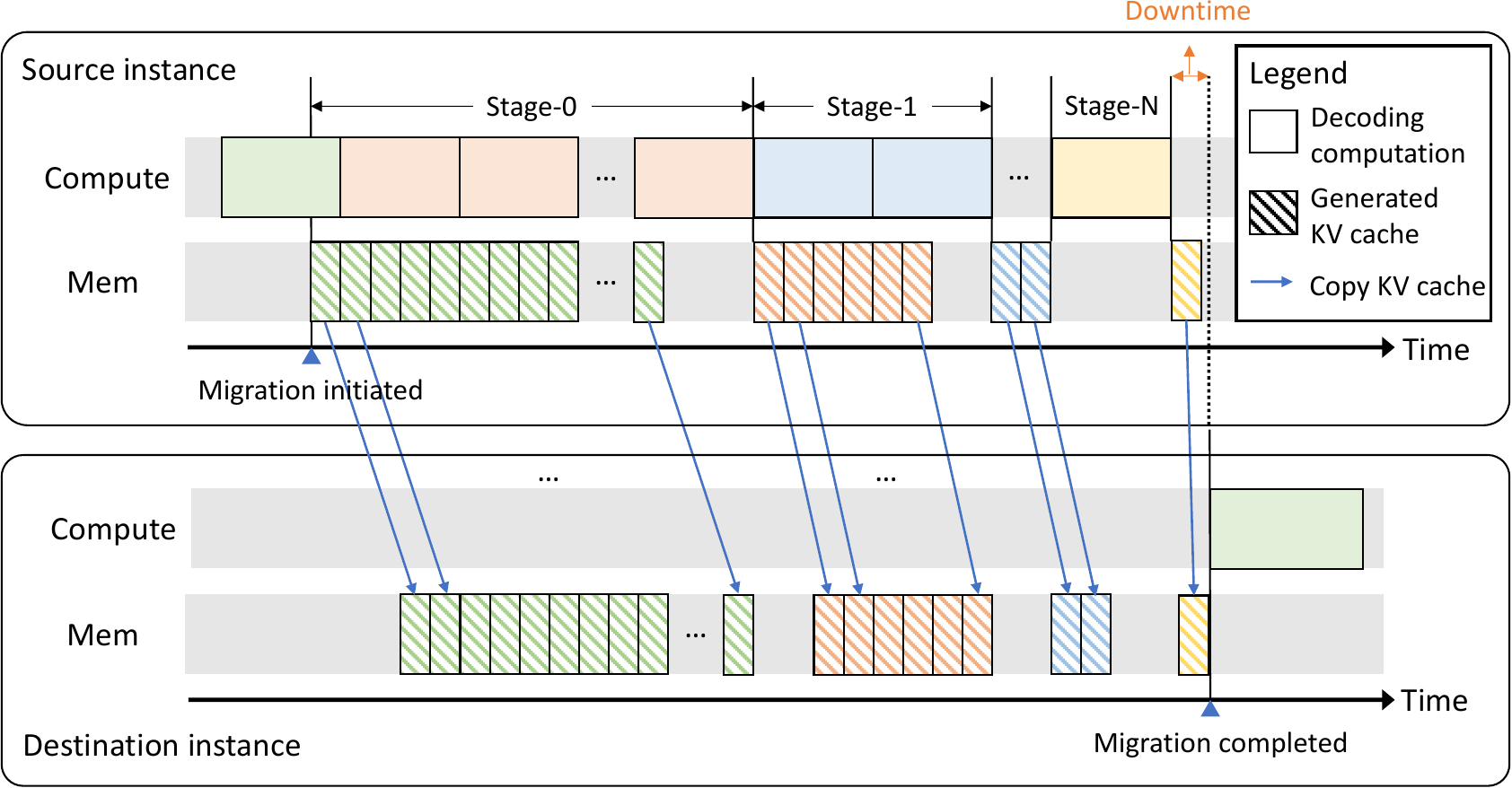}
    \caption{\sysname{} adopts multi-stage migration to overlap the computation and KV cache copying for minimal downtime.}
    \label{fig:design-migration}
\end{figure}

The live migration mechanism of \sysname{} utilizes the inherent append-only characteristic of KV cache to pipeline the KV cache copying with the decoding computation. Because the KV cache already generated won't be modified in the following iterations, \sysname{} can safely copy the KV cache of previous tokens in parallel with the computation for new tokens. In this way, \sysname{} achieves \emph{near-zero} and \emph{constant} downtime to the rescheduled request.
As shown in Figure \ref{fig:design-migration}, when migration is initiated, the source instance starts to copy the KV cache blocks of completed iterations, and continues the computation at the same time (stage 0). When the copying for the previous KV cache blocks is done, there will be a few more iterations (\ie, blocks in Figure~\ref{fig:design-migration}) computed in stage 0. Then, it switches to stage 1 to copy the KV cache generated by stage 0, while continuing the computation afterwards. The copying is generally much faster than the computation, thus the number of new blocks is typically small such that we can copy them in a very short period.
To the end, only one iteration of computation is conducted for the KV cache migration (\ie, stage-N).
Therefore, \sysname{} suspends the computation for the request by draining it out of the current batch and copies the remaining block, which introduces the downtime of this request.
Once it is finished, the migration completes
and the request resumes on the destination instance.
Although the total copying duration of the whole sequence depends on the sequence length, the downtime for the request is only the period of copying the KV cache generated by one iteration, which is negligible regardless of the sequence length.

\begin{figure}[t]
    \centering
    \includegraphics[width=0.99\columnwidth]{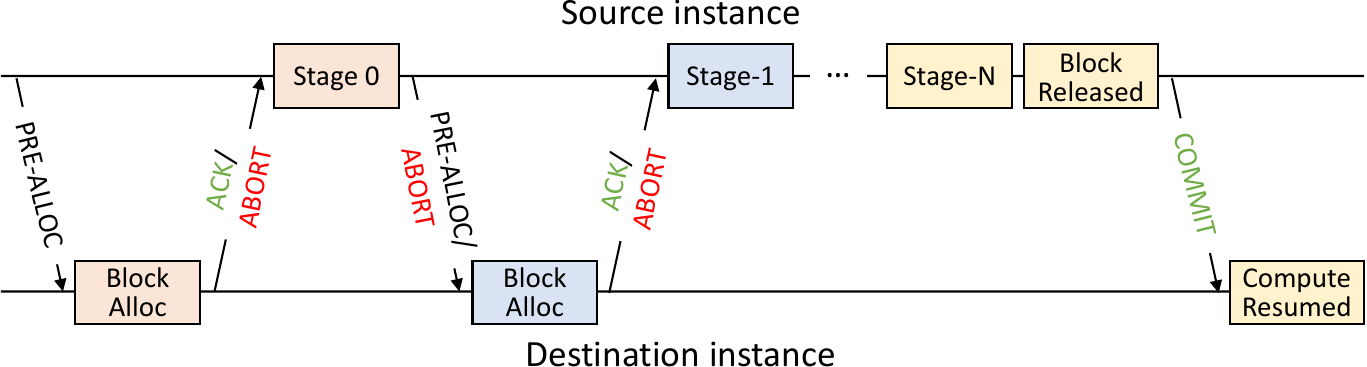}
    \caption{\revise{Handshake during migration.}}
    \label{fig:handshake}
    \vspace{-10pt}
\end{figure}

The request migration approach of \sysname{} borrows the key concept introduced in virtual machine (VM) live migration~\cite{vmmigration2005}, which gradually reduces the working set to minimize the downtime. \sysname{} does not require the dirty page tracing in VM migration as the working set (\ie, KV cache) is append-only and does not change during migration.
However, LLM serving further introduces additional challenges.
Firstly, as both the source and destination instances are continually processing requests, the request might run out of memory during migration. 
Secondly, the request can complete in the middle of migration, due to the unpredictable execution (\ie, generating EOS token) and the continuous batching~\cite{orca2022}.
To handle such exceptions and guarantee correctness during the asynchronous computation and memory copying, \sysname{} introduces fine-grained coordination between the participating instances with a \emph{handshake} process (Figure \ref{fig:handshake}).
Before each stage, the source instance issues a pre-allocate request with the number of blocks to migrate to make sure that the destination has enough space. The destination will try to allocate and reserve the blocks; if it succeeds or fails, the destination will notify the source to proceed or abort the migration and clean the states, respectively. Similarly, after each stage, the source instance also checks whether the request being migrated has completed or been preempted --- if it has, the source will notify the destination to abort and release the reserved blocks; otherwise the source will go ahead to the next stage. 
\revise{The source or destination will also abort the migration if the other side fails.}
After the final stage finishes, the source releases its local blocks and notifies the destination to commit the migration and resume the execution of the request.

\subsection{Distributed Scheduling Architecture}\label{sec:arch}

\revise{
The live migration mechanism provides the foundation for runtime rescheduling of LLM inference requests. However, achieving fully dynamic scheduling is still non-trivial due to the higher scheduling pressure than in traditional schedulers.
In particular,
\sysname{} would need to continuously track and reschedule every single running request throughout the cluster, rather than only dispatch incoming requests for one time or only manage running requests on one instance. This implies a higher scheduling frequency and a larger number of requests for the scheduler to track and schedule in each round.
}

\begin{figure}[t]
    \centering
    \includegraphics[width=0.8\columnwidth]{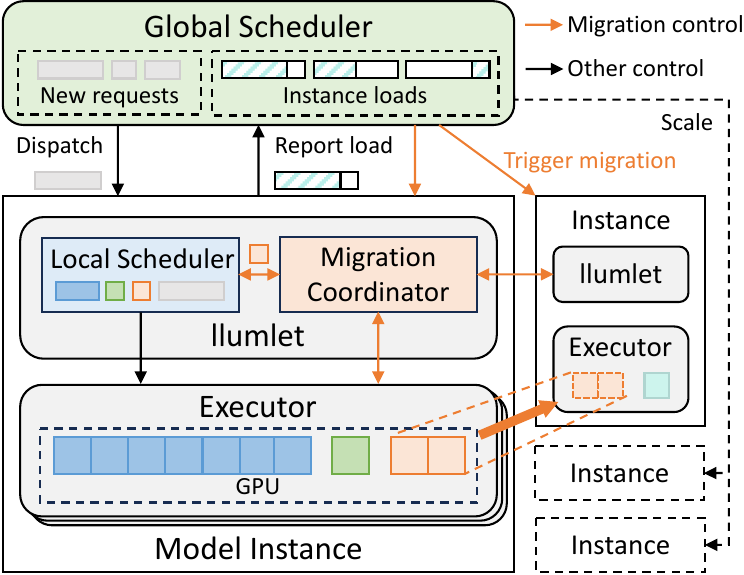}
    \caption{\sysname{} architecture.}
    \label{fig:arch}
\end{figure}

\sysname{} devises a scalable architecture that combines a cluster-level \emph{global scheduler} and distributed instance-level schedulers, named \emph{\syslet{}}s, to enable continuous rescheduling efficiently (Figure \ref{fig:arch}). \sysname{} defines a clean separation of concerns with a narrow interface between the two levels. \revise{The global scheduler does not directly track or schedule the running requests; instead,} it makes all scheduling decisions oriented to the \emph{instances}, according to the \emph{memory loads} of them. This way, \revise{the complexity of the global scheduler remains independent from the running requests, thereby preserving similar scalability to schedulers without dynamic scheduling.}
The loads are reported by the \syslet{}s periodically, based on the request status and \sysname{}'s scheduling policy.

The global scheduler utilizes the load information to dispatch new requests, trigger migration across instances, and control the instance auto-scaling. 
In particular, for migration, the decisions are not made for specific requests; the global scheduler just pairs the source and destination instances, only based on the loads, and marks them as in the corresponding states to trigger the migration. The \syslet{}s will decide the requests to migrate and execute the migration automatically.

The \syslet{} of each instance consists of a local scheduler and a migration coordinator. In addition to the functionalities of similar roles in existing systems like queuing, batching, and block management, an important new task of the local scheduler is to calculate the memory load of the instance. The load is not simply the physical memory being used; instead, it is  a sum of the ``virtual usages'' (\S\ref{sec:policy}) of the requests.
    The local scheduler is also responsible for deciding the requests to migrate when triggered.
Given the chosen requests, the migration coordinator will coordinate with the local scheduler and the other instance, and instruct the model executor to do the memory copying, as described before.

\subsection{Dynamic Scheduling Policy}\label{sec:policy}

\subsubsection{Goals and Definitions}

\sysname{}'s scheduling policy is designed with the following goals. The first is to improve \textbf{prefill and decode latencies}, by reducing queuing delays, preemptions, and interference.
The second goal is \textbf{load-adaptivity} to handle varying cluster load and improve cost efficiency. We notice that the benefits of rescheduling is also relevant to cluster load, which could be limited under too high/low load. \sysname{} incorporates instance auto-scaling to keep appropriate cluster load for both saving costs and maximizing the benefits of rescheduling.

Besides these two goals similar to those of existing systems, \sysname{} introduces a new goal of request \textbf{priorities} that comes from the new requirements of LLMs. Priorities present a systematic approach for the same LLM to serve certain requests with higher emergency, \eg, from ChatGPT Plus or more interactive applications.
\sysname{} provides applications with an interface for specifying request priorities to meet different SLOs, in terms of \emph{scheduling priority} and \emph{execution priority}.
Requests with higher scheduling priorities will get scheduled earlier to reduce their queuing delays. Those with higher execution priorities will be given lower instance load and hence less interference to accelerate their execution. Currently, \sysname{} supports two priority classes, high and normal, to demonstrate the ability of \sysname{} to prefer high-priority requests, but our design also generalizes to more priorities.

\subsubsection{Virtual Usage}

\revise{To achieve the multiple goals above under the distributed scheduling architecture, \sysname{} needs a scheduling policy that can express these goals using simple instance-level metrics, to improve the efficiency and scalability of the global scheduler.
To this end, \sysname{} introduces the \emph{virtual usage} abstraction to unify these different, sometimes conflicting goals into a simple load metric of instances. 
The key observation here is that the aforementioned rescheduling scenarios fall into two categories: \emph{load balancing}, and \emph{creating free space on one instance} (de-fragmentation, prioritization, and draining out instances).}
We find that they can be unified into load balancing by assuming a virtual load on the instance: to create free space on an instance, we just need to set the virtual usages of certain requests to make the instance virtually overloaded, then a load balancing policy will be triggered to migrate the requests to other instances.

\begin{figure}[t]
    \centering
    \includegraphics[width=0.9\columnwidth]{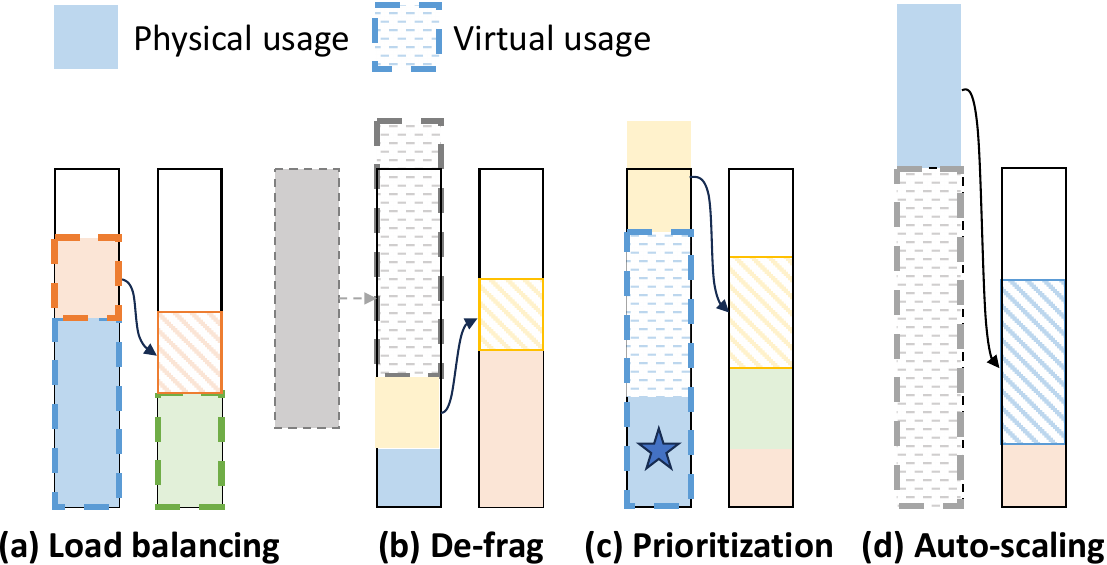}
    \caption{\sysname{} combines virtual usages with a load-balancing policy to unify multiple scheduling goals.}
    \label{fig:vmu}
    \vspace{-10pt}
\end{figure}

This observation leads us to a simple heuristic with load-balancing as the basis, combined with a set of rules for setting request virtual usages in different situations. We summarize the rules in the function \texttt{CalcVirtualUsage} in Algorithm \ref{algorithm:policy} and illustrate example scenarios in Figure \ref{fig:vmu}.
In normal cases, the virtual usage of a request is just its physical memory usage to enable routine load balancing, as shown in Figure~\ref{fig:vmu}(a). We discuss the rules for other cases as follows.

\paragraph{Queuing requests.}
For the head-of-line queuing request on an instance, we assign a positive virtual usage to it to reflect its resource demand in terms of the required memory, although the physical usage is 0. Thus, queuing requests will increase the total virtual usage of the instance, then the policy will trigger migration for load balancing (which in effect is de-fragmentation for the queuing request), as shown in Figure~\ref{fig:vmu}(b). There could be a lot of heuristics to explore for setting the virtual usage, which controls the tradeoff between reducing queuing delays and load balancing --- for example, gradually increasing the virtual usage of a queuing request until it reaches the real memory demand. \sysname{} currently uses a simple rule that directly uses its real demand (line \ref{line:queuing} in Algorithm \ref{algorithm:policy}), which favours reducing queuing delays. This rule is based on our observation that queuing delay can dominate the end-to-end latency and worth such preference. Our evaluation also shows that this rule preserves the benefits of load balancing, due to the high flexibility of migration.

\paragraph{\revise{Execution} priorities.}
For a request with high execution priorities, \sysname{} tries to prevent the instance the request is running on from exceeding a given level of real load, by reserving a memory space as headroom, as shown in Figure~\ref{fig:vmu}(c). This is achieved by adding such a headroom on the physical usage of a high-priority request to get the virtual usage (line \ref{line:priority}). When there are multiple high-priority requests on an instance, this headroom is divided among them (line \ref{line:headroom}). The headroom for high-priority requests is currently defined as that required to preserve the ideal decode speed (\ie, no visible interference), which is obtained through profiling. The headroom for normal requests is 0. \sysname{} can also support more execution priorities by specifying the sizes for the headroom.
When the headroom for a high-priority request is running up, the other normal requests will be migrated away by the load balancing policy because the instance is overloaded in terms of the total virtual usage. 

\paragraph{Auto-scaling.} When a new instance is launched, \sysname{}'s load balancing policy will automatically saturate it by migrating requests from other instances to it. When an instance is terminating, we artificially add a fake request with a virtual usage of infinity on it (line \ref{line:infinity}), then the remaining requests will be migrated to other instances, as shown in Figure~\ref{fig:vmu}(d).

\subsubsection{Policies}

{
\IncMargin{0.2em}
\begin{algorithm}[bt]
\SetKwData{Left}{left}\SetKwData{This}{this}\SetKwData{Up}{up} 
\scriptsize
\DontPrintSemicolon
\SetKwFunction{CalcVirtualUsage}{CalcVirtualUsage} \SetKwInOut{Input}
{Union}\SetKwFunction{GetHeadroom}{GetHeadroom} \SetKwInOut{Input}
{Union}\SetKwFunction{AddFakeReq}{AddFakeReq} \SetKwInOut{Input}
{Union}\SetKwFunction{CalcFreeness}{CalcFreeness} \SetKwInOut{Input}
{input}\SetKwInOut{Output}{output}
\SetKw{Break}{break}
\SetKwProg{Fn}{Function}{:}{}

\Fn{\CalcVirtualUsage{$req$, $instance$}}{
\If{$req.isQueuing$}{
\If{$req.isHeadOfLine$}{
\KwRet $req.demand$\;\label{line:queuing}
}
\KwRet 0\;
}
\If{$req.isFake$}{
\KwRet $\infty$\;\label{line:infinity}
}
\KwRet $req.physicalUsage+$\GetHeadroom{$req.priority,instance$}\;\label{line:priority}
}

\Fn{\GetHeadroom{$p$, $instance$}}{
\KwRet $headroomForPriority[p]/instance.numRequests[p]$\;\label{line:headroom}
}

\Fn{\CalcFreeness{$instance$}}{
\If{$instance.isTerminating$}{
\AddFakeReq($instance.requests$)\;
}$totalVirtualUsages=0$\;
\For{$req$ in $instance.requests$}{
$totalVirtualUsages+=$\CalcVirtualUsage{$req,instance$}\;
}
$freeness=(instance.M-totalVirtualUsage)/instance.B$\;

\KwRet $freeness$\;
}

\caption{Virtual Usage and Freeness Calculation}
\label{algorithm:policy}
\end{algorithm}
\DecMargin{0.2em} 
}
We then describe how the specific scheduling decisions are made based on the virtual usages.

\paragraph{Dispatching.}
\revise{\sysname{} dispatches new requests with higher scheduling priorities first. Within the same priority, it adopts a simple first-come-first-serve order. On each instance, requests are scheduled in the same order.}
\sysname{} uses a load-balancing policy that dispatches each request to the freest instance. We introduce a metric for measuring the \emph{freeness} of an instance defined as $F=(M-\sum{V})/B$, where $M$ is the total memory, $V$ is the virtual usage of each request, and $B$ is the batch size. While $(M-\sum{V})$ already measures the free space, we divide it by the batch size because it determines the consumption speed, \ie, the number of new tokens per iteration. Thus the metric suggests how many iterations the batch can still run for. Then \sysname{} dispatches each incoming request to the instance with the highest freeness.
Because the virtual usage of a request can be larger than the physical, it is possible that $F$ is a negative value, \eg, when there are queuing requests or high-priority requests. Such negative freeness values help \sysname{} automatically treat such instances as overloaded and prefer dispatching requests to other instances. The freeness metric also guides the migration and auto-scaling, as shown later.



\paragraph{Migration.}
\sysname{} triggers the migration policy periodically. In each round, \sysname{} selects the candidate sets of source and destination instances by choosing those with freeness values smaller or greater than given thresholds, respectively. \sysname{} pairs the instances from both sets by picking the two with the lowest and the highest freeness values repeatedly, and then sets them in corresponding states.
The \syslet{} of each source instance then starts to migrate requests to the destination continuously, until it is no longer set in the source state. The \syslet{} prefers the requests with lower priorities and shorter sequence lengths when choosing the requests to migrate. In the next round, if an instance during migration is no longer beyond the thresholds, \sysname{} will unset the migration state and the migration will stop. 

\paragraph{Auto-scaling.} \sysname{} scales the instances according to the cluster load in terms of the averages freeness for the normal priority across instances. The policy maintains the average freeness within a range $[x, y]$, and adds or terminates an instance when the freeness is smaller than $x$ or greater than $y$ for a period, respectively. \sysname{} chooses the instance with fewest running requests for termination.

\section{Implementation}\label{sec:impl}

\revise{We implement \sysname{} with 3,300 lines of Python code. \sysname{} is a standalone library comprising both its own components and an interface to integrate and communicate with backend inference engines. This architecture makes \sysname{} non-intrusive and extensible to different backends. \sysname{} currently supports vLLM~\cite{vllmcode} as the backend,}
which is an open-source state-of-the-art inference engine that features continuous batching, PagedAttention, and tensor-parallel distributed inference~\cite{vllm2023,megatron2019}.

\paragraph{Multi-instance serving.}
\revise{
\sysname{} instantiates the multiple instances of the backend and the other components as Ray~\cite{ray18} actors.
Ray's Python-native distributed runtime enables fine-grained coordination among these actors in a simple and efficient manner.
\sysname{} also launches a set of request frontend actors that exposes an OpenAI-style API endpoint~\cite{openaiapi}. Although a request can be migrated across backend instances, the generated tokens are forwarded to the frontend and then returned to end users, ensuring a steady API service.
}

\paragraph{KV cache transfer.}
We use the Gloo collective communication library~\cite{gloo} (the \texttt{Send}/\texttt{Recv} primitives) for the KV cache transfer during migration. A potential alternative is NCCL~\cite{nccl}, which is generally faster than Gloo on GPUs but has been adopted in communication for distributed inference. However, \sysname{} needs to migrate requests in parallel with the inference to minimize the downtimes, but concurrent invocations of NCCL are known to be unsafe~\cite{ncclconcurrency}. The pipelined migration design allows us to use Gloo while maintaining negligible downtimes.
\revise{Using Gloo needs to copy the KV cache between CPU and GPU memory, which is done in another CUDA stream to avoid blocking the inference computation.}
Note that in typical deployments, the communication-heavy tensor parallelism is limited in a single machine for high-speed transfer~\cite{megatron2021}. In such cases, migration between instances (machines) will not interfere with the tensor-parallel inference.

\paragraph{Block fusion.}
vLLM stores the KV cache in non-contiguous small blocks that are dynamically allocated. For example, the block size of a 16-bit LLaMA-7B model is 128 KB (for key or value tensors of 16 tokens in each layer), and a sequence of 1k tokens translates to 4k such blocks (32 layers). \revise{
To avoid the overhead of sending these blocks using many small messages,
we fuse the blocks by copying them from GPU memory to a contiguous CPU memory buffer and use Gloo to send the buffer as a whole, thereby improving the transfer efficiency.}

\paragraph{Fault tolerance.}
\revise{
\sysname{} provides fault tolerance for each component to ensure high service availability.
When the global scheduler fails, \sysname{} temporarily falls back to a scheduler-bypassing mode, thus not affecting the service availability:
that is, the request frontends directly dispatch requests to certain instances using simple rules, and migration is disabled. When an instance (or the co-located \syslet{}) fails, the requests running on it will be aborted. In particular, ongoing migration on failed instances will also be aborted (the request being migrated is not necessarily aborted, depending on if its source instance is healthy), which is handled by the handshake process. These failed actors will be automatically restarted by Ray, after which the service could go back to normal state.
}

\section{Evaluation}\label{sec:eval}

We evaluate \sysname{} on a 16-GPU cluster using realistic models and various workloads. Overall, our key findings include:

\begin{packed_itemize}
    \item \revise{\sysname{} introduces near-zero downtime to requests being migrated and near-zero overhead to other running requests.}
    \item \sysname{} improves prefill latencies by up to 15$\times$/7.7$\times$ (P99/mean) over INFaaS on 16 LLaMA-7B instances via de-fragmentation. \sysname{} also improves P99 decode latency by up to 2$\times$ by reducing preemptions.
    \item \sysname{} improves high-priority request latencies by up to 1.5$\times$ by reducing their queuing delays and accelerating their execution, while preserving similar performance of the normal requests.
    \item \sysname{} achieves up to 36\% cost saving while preserving similar P99 latencies with efficient auto-scaling.
\end{packed_itemize}

\subsection{Experimental Setup}

\paragraph{Testbed.}
We use a 16-GPU cluster with 4 GPU VMs on Alibaba Cloud (type \texttt{ecs.gn7i-c32g1.32xlarge}), each with 4 NVIDIA A10 (24 GB) GPUs connected via PCI-e 4.0, 128 vCPUs, 752 GB memory, and 64 Gb/s network bandwidth.

\paragraph{Models.}
We conduct experiments using a popular model family, LLaMA~\cite{touvron2023llama}.
We test two different specifications: LLaMA-7B, which runs on a single GPU, and LLaMA-30B, which runs on 4 GPUs of a machine using tensor parallelism.
\revise{The models adopt the commonly used 16-bit precision.}
The version of vLLM that we based on only supports the original LLaMA with a maximum sequence length of 2k, but there have been a series of recent LLaMA variants supporting longer sequence lengths ranging from 4k to 256k~\cite{longllama,codellama,xiong2023effective,touvron2023llama2}. Since the model architectures and inference performance of these variants are mostly similar to those of LLaMA, we believe that our results are representative of more model types and larger sequence length ranges from a systems perspective.

\paragraph{Traces.}
Similar to prior work~\cite{orca2022,vllm2023,alphaserve2023}, we synthesize request traces to asses \sysname{}'s online serving performance.
We use Poisson and Gamma distributions with different request rates (requests per second) to generate request arrivals. For Gamma, we also use varying coefficients of variance (CVs) to adjust the burstiness of the requests.
Each trace has 10,000 requests. 
We choose an appropriate range of request rates or CVs for the traces to maintain the loads within a reasonable range: nearly no queuing delays and preemptions for P50 requests, and queuing delays within a few tens of seconds for P99 requests when using \sysname{}.

For the input/output lengths of requests, \revise{we use two public ChatGPT-4 conversation datasets, ShareGPT (GPT4)~\cite{sharegpt4} and BurstGPT (GPT4-Conversation)~\cite{wang2024efficient}, for an evaluation on real workloads.
Considering that \sysname{} targets more diversified applications, we also use generated power-law length distributions} to emulate long-tail workloads that mix both frequent, short sequences (\eg, for interactive applications like chatbots and personal assistants) and seldom, long sequences (\eg, summarizing or writing articles). We generate multiple distributions with different long-tail degrees and mean lengths ($128$, $256$, $512$), as shown by the Short (S), Medium (M), and Long (L) distributions in Table~\ref{tab:dist}. These distributions have a maximum length of 6k, thus the total sequence length of a request (input plus output) will not exceed the capacity of an A10 GPU when running LLaMA-7B (13,616 tokens).
To observe the performance with different workload characteristics, we construct the traces by picking different combinations of the length distributions for inputs and outputs as follows: S-S, M-M, L-L, S-L, and L-S.

\begin{table}[t]
\footnotesize
\centering
\begin{tabular}{c|c|c|ccccc}
\toprule
\multicolumn{3}{c|}{\textbf{Distribution}}                   & \textbf{Mean} & \textbf{P50} & \textbf{P80} & \textbf{P95} & \textbf{P99} \\ \midrule
\multirow{4}{*}{Real} & \multirow{2}{*}{ShareGPT} & In  & 306             & 74            & 348              & 1484               & 3388            \\
                      &                           & Out & 500             & 487            & 781              & 988               & 1234            \\ \cmidrule{2-8}
                      & \multirow{2}{*}{BurstGPT} & In  & 830             & 582            & 1427              & 2345               & 3549            \\
                      &                           & Out & 271             & 243            & 434              & 669               & 964            \\ \midrule
\multirow{3}{*}{Gen}  & \multicolumn{2}{c|}{Short (S)}           & 128           & 38           & 113            & 413             & 1464         \\
                      & \multicolumn{2}{c|}{Medium (M)}          & 256           & 32           & 173            & 1288            & 4208         \\
                      & \multicolumn{2}{c|}{Long (L)}            & 512           & 55           & 582            & 3113            & 5166        \\ \bottomrule
\end{tabular}
\caption{Real and generated distributions of sequence lengths (numbers of tokens) used in our evaluation. The real distributions include those of both inputs (``In'') and outputs (``Out'').}\label{tab:dist}
\end{table}

\paragraph{Baselines.}
We compare \sysname{} with the following schedulers. 
All the baselines and \sysname{} use vLLM as the underlying inference engine to focus the comparison on the request scheduling across instances. 



\begin{packed_itemize}
    \item \emph{Round-robin dispatching}: a simple dispatching policy to distribute requests across instances evenly, which is a typical behavior of production-grade serving systems~\cite{triton,rayserve,deepspeedmii}.
    \item \emph{INFaaS++}: an optimized version of INFaaS~\cite{romero2021infaas}, a state-of-the-art scheduler for multi-instance serving. We evaluate its load-balancing dispatching and load-aware auto-scaling policies. We improve it by making it focus on the GPU memory load as it is the dominant resource in LLM serving. This load also counts in the memory required by queuing requests on each instance to reflect the queue pressure.
    \item \emph{\sysname{}-base}: a base version of \sysname{} that is priority-agnostic (\ie, treats all requests as the same priority) but enables all the other features including migration.
\end{packed_itemize}

\paragraph{Key metrics.}
We focus on request latency, in terms of end-to-end, prefill (that of the first generated token), and decode (that since first generated token to the last, averaged over all generated tokens). 
We report both mean and P99 values.

\begin{figure}[t]
    \centering
    \includegraphics[width=0.99\columnwidth]{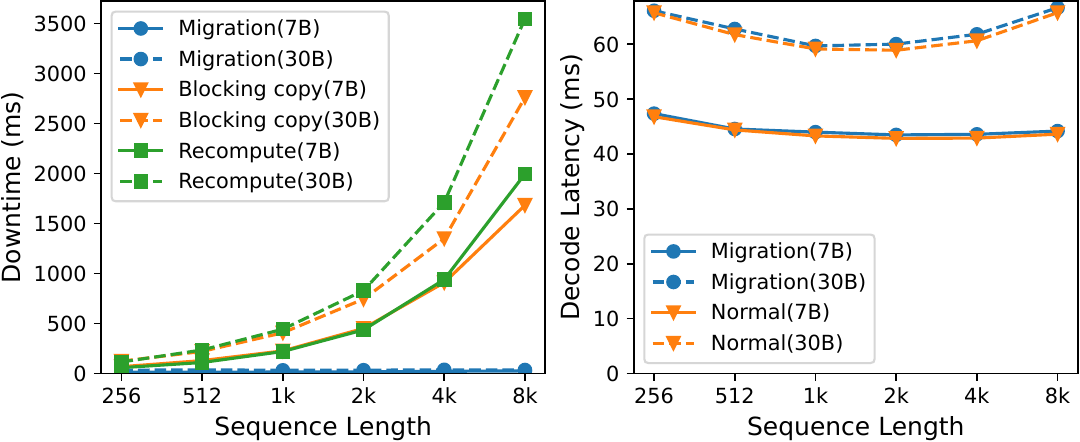}
    \caption{Downtime and overhead of migration.}
    \label{fig:microbench}
\end{figure}

\begin{figure*}[t]
    \centering
    \includegraphics[width=0.95\textwidth]{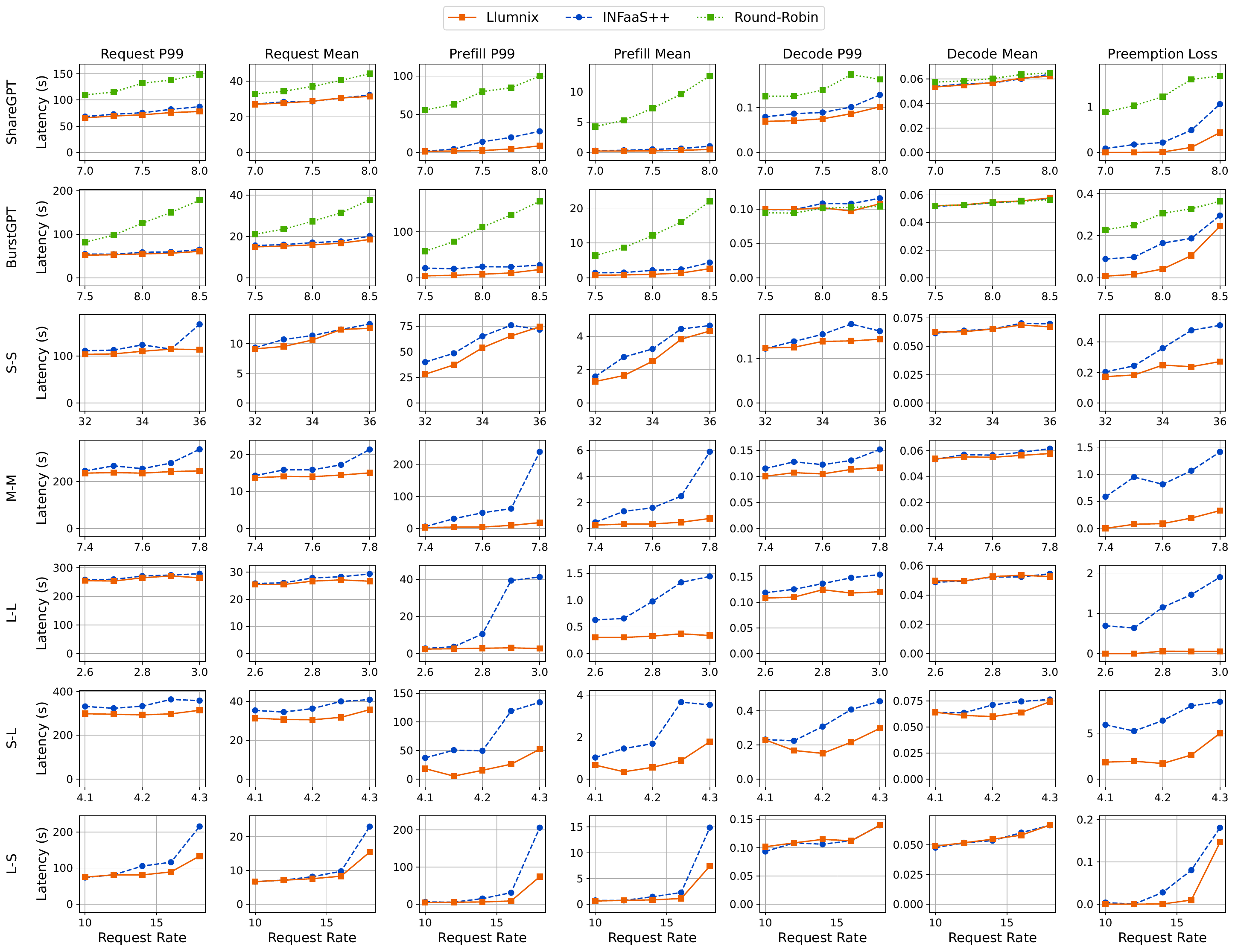}
    \caption{Request end-to-end, prefill, and decode latencies and preemption loss of serving 16 LLaMA-7B instances. Each row shows a set of experiments using a trace with a specific sequence length distribution, as annotated on the Y-axis labels.}
    \label{fig:eval-serving}
\end{figure*}


\subsection{Migration Efficiency}
\label{eval:migrate}

We first examine the performance of \sysname{}'s migration mechanism, in terms of the downtimes introduced to the migrated requests and the performance overheads for the running requests. We test both the 1-GPU LLaMA-7B and the 4-GPU LLaMA-30B models. For each model, we deploy two instances on two different machines. We use different sequence lengths, for each of which we run a batch of requests with the same total length of 8k on both instances.
We migrate one of the requests from one instance to another and measure its downtime and \revise{the decode speeds of the running batches on both instances during migration.}

We compare the downtime during migration with two simple approaches: recomputing, and blocking copying of the KV cache using Gloo (non-blocking for other requests).
As shown in Figure~\ref{fig:microbench} (left), the downtime of migration is nearly constant with increasing sequence lengths (roughly 20-30 ms), even shorter than a single decode step. In comparison, the downtimes of baselines increase with the sequence lengths, reaching up to \revise{111$\times$ that of migration. For example, recomputing an 8k sequence for LLaMA-30B takes 3.5s, which translates to a service stall similar to 54 decode steps}. We also notice that for all sequence lengths, the migration only takes two stages, which is the minimum. This is because the data copying is sufficiently fast and the number of new tokens generated during the first stage is small.

Figure~\ref{fig:microbench} (right) also compares the per-step decode times during migration \revise{on the source instance with that during normal execution (results on the destination are mostly similar). We observe up to 1\% performance differences for both LLaMA-7B and LLaMA-30B, showing the negligible migration overhead.
Also note that such overhead exists only when there are requests being migrated (in or out) on an instance. We find that in all the serving experiments in the following sections, the average fraction of time span with ongoing migration for each instance is only roughly 10\%. This implies an effective overhead that is even much smaller, which is worthwhile for the great scheduling benefits of migration.}
\begin{figure}
    \centering
    \includegraphics[width=0.8\columnwidth]{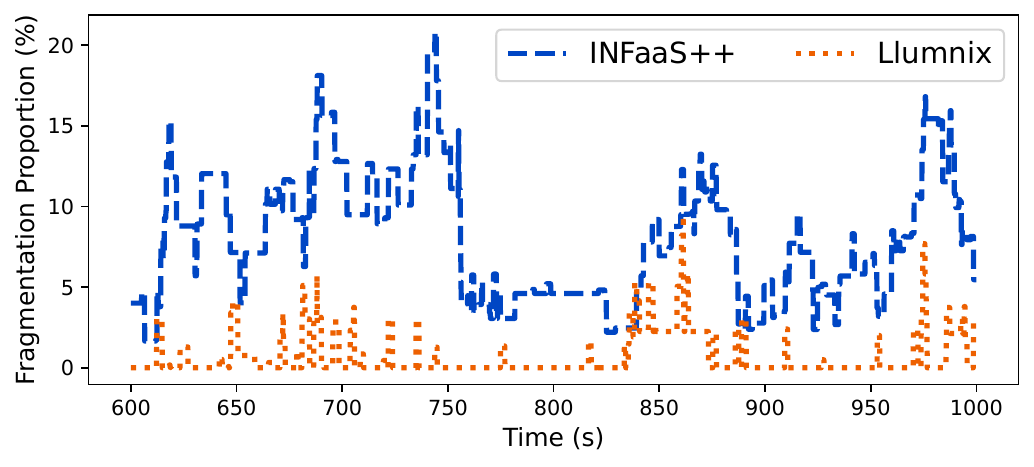}
    \caption{Memory fragmentation over time.}
    \label{fig:eval-frag}
\end{figure}

\begin{figure*}[t]
    \centering
    \includegraphics[width=0.95\textwidth]{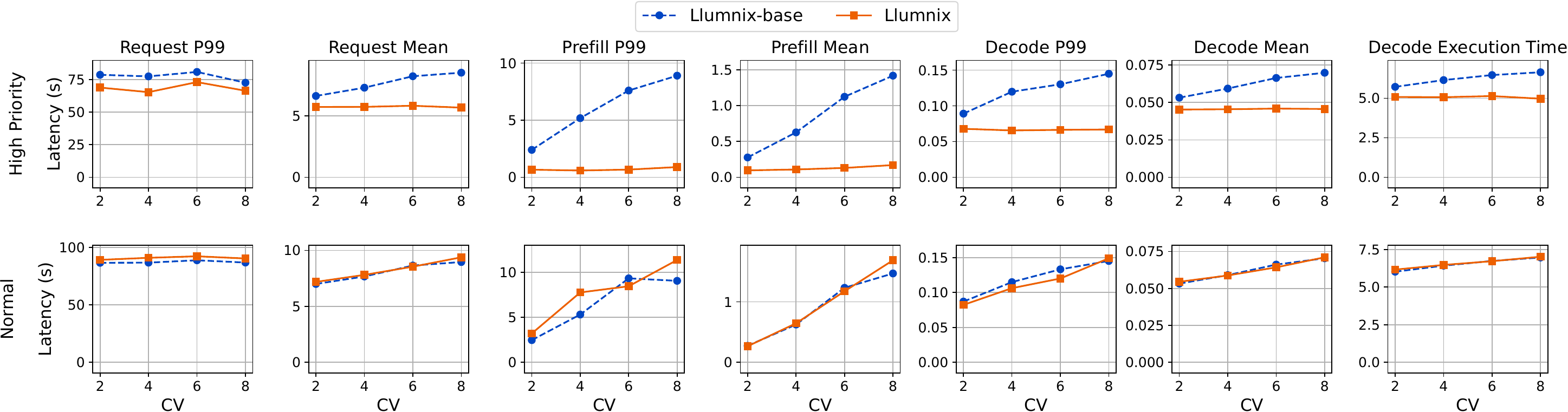}
    \caption{Performance of high-priority and normal requests, as annotated on the Y-axis labels.}
    \label{fig:eval-priority}
\end{figure*}

\subsection{Serving Performance}

We evaluate the scheduling performance of \sysname{} in online serving using 16 LLaMA-7B instances (auto-scaling is disabled except in experiments in \S\ref{sec:auto-scaling}).


\paragraph{Real datasets.}
\revise{
We first compare \sysname{} with round-robin and INFaaS++ using the ShareGPT and BurstGPT traces (the top two rows in Figure \ref{fig:eval-serving}).
\sysname{} outperforms the baselines in end-to-end request latency by up to 2$\times$ and 2.9$\times$ for mean and P99, respectively. In particular, we observe that round-robin always performs much worse than both INFaaS++ and \sysname{}: since the sequence lengths have high variance, simply distributing requests evenly can still lead to unbalanced load, impacting both prefill and decode latencies.
\sysname{} achieves significant gains in prefill latency over round-robin, by up to 26.6$\times$ for mean and 34.4$\times$ for P99. This is because round-robin can possibly dispatch new requests to overloaded instances, leading to long queuing delays. \sysname{} also improves P99 decode latency by up to 2$\times$, by load balancing to reduce preemptions. This margin seems smaller as the latency penalty caused by preemptions is averaged over all generated tokens. However, whenever preemption occurs, it results in a sudden service stall, which impacts user experience.
Figure \ref{fig:eval-serving} (the rightmost column) reports the preemption loss in terms of the extra queuing and recomputing times (mean value of all requests).
\sysname{} reduces preemption loss by 84\% on average compared to round-robin. These results highlight the importance of load balancing in LLM serving. In the following experiments using generated distributions with higher variance, round-robin showed up to two orders of magnitude worse latencies.
Therefore, we omit it for the other traces for clarity of the figures and focus on the comparison between INFaaS++ and \sysname{}.
}

\final{\sysname{} outperforms INFaaS++ in mean and P99 prefill latencies by up to 2.2$\times$ and 5.5$\times$, and P99 decode latencies by up to 1.3$\times$, respectively, showing the extra benefits of migration, beyond dispatch-time load balancing. Next we use more traces with different characteristics to further evaluate them for a deeper understanding of the improvements.}


\paragraph{Generated distributions.}
\revise{We compare \sysname{} and INFaaS++ using multiple generated distributions
(bottom five rows in Figure \ref{fig:eval-serving}).}
\sysname{} outperforms INFaaS++ across all traces in end-to-end request latency by up to 1.5$\times$ and 1.6$\times$ for mean and P99, respectively. 
For prefill, the improvements are up to 7.7$\times$ for mean and 14.8$\times$ for P99.
Despite dispatching requests to instances with the lowest load, INFaaS++ can still exhibit long queuing delays due to fragmentation, especially for the long-tail requests with long inputs.
\sysname{} uses migration for de-fragmentation to reduce such queuing delays, showing more gains in traces with more long inputs.

To take a closer look at the memory fragmentation, we further present a case study on the experiment of the M-M trace with the request rate of 7.5. We define the fragmented memory at each moment as the portion of cluster free memory that could satisfy the demands of the head-of-line blocking requests across all instances, if no fragmentation. For example, if the total free memory is 8 GB, with three head-of-line blocking requests each requiring 3 GB, then the fragmented memory is counted as 6 GB, \ie, this 6 GB memory could satisfy two queuing requests if no fragmentation. This metric suggests the memory space wasted due to fragmentation. We report the proportion of fragmented memory in the cluster total memory. In the example, if the total memory is 16 GB, then the proportion is 37.5\% (6/16).
Figure \ref{fig:eval-frag} shows the fragmentation proportion of the experiment during a busy period. We observe that INFaaS++ often shows higher than 10\% fragmentation, wasting a significant amount of cluster memory. In comparison, the fragmentation is often 0 in \sysname{}. The average values during this period are 0.7\% and 7.9\% for \sysname{} and INFaaS++ respectively (92\% reduction),
highlighting the effect of de-fragmentation using migration.

\sysname{} also improves the P99 decode latency by up to 2$\times$, through migration to reduce preemptions. Although INFaaS++ already implements load balancing in dispatching to reduce preemptions, migration complements it by reacting to the real sequence lengths, which are unknown at request arrivals.
As shown in Figure \ref{fig:eval-serving}, \sysname{} significantly reduces the preemption loss, in many cases down to near zero. The reduction is 70.4\% on average across all experiments, which translates to an average reduction of 1.3 seconds in the end-to-end request latency.



\subsection{Support for Priorities}

We evaluate the support for priorities of \sysname{} by randomly picking 10\% of the requests and assigning high scheduling and execution priorities. We use traces with the Short-Short length distribution and Gamma arrival distribution. We vary the CV parameter  to show the interference to high-priority requests due to bursty workloads and load spikes. We empirically choose a target memory load of 1,600 tokens for high-priority requests, as we observe that such load preserves near-ideal decode speed (refer to Figure~\ref{fig:interference}). \sysname{} translates this target load to the corresponding memory headroom for high-priority requests. We compare \sysname{} with \sysname{}-base, which simply treats all requests as the same priority.

As shown in the first row in Figure~\ref{fig:eval-priority}, \sysname{} improves mean request latencies for the high-priority by 1.2$\times$ to 1.5$\times$ with increasing CVs. Higher CVs leads to more high-load periods, where high-priority requests can suffer more interference if not protected. Even with higher CVs, \sysname{} still delivers similar latencies of high-priority requests, showing the isolation \sysname{} provides to such requests. This is because \sysname{} can handle changing high-priority loads by dynamically creating space for them, which is difficult in approaches like static resource reservation. For prefill latencies, \sysname{} shows 2.9$\times$ to 8.6$\times$ gains for the mean, and 3.6$\times$ to 10$\times$ for the P99, respectively. This is achieved by reducing the queuing delays with high scheduling priorities. \sysname{} also improves decode latencies by 1.2$\times$ to 1.5$\times$ for the mean and 1.3$\times$ to 2.2$\times$ for the P99, respectively. This improvement comes from the acceleration of the decode computation by giving lower instance loads and interference to high execution priorities, shown by the similar gains in the average decode computation time (the rightmost column).
We also notice that \sysname{} preserves similar performance of the normal requests (the second row in Figure \ref{fig:eval-priority}): \sysname{} increases the mean request, prefill, and decode latencies of normal requests by up to 4.5\%, 13\%, and 2\%, respectively. 

\begin{figure*}[t]
    \centering
    \includegraphics[width=0.95\textwidth]{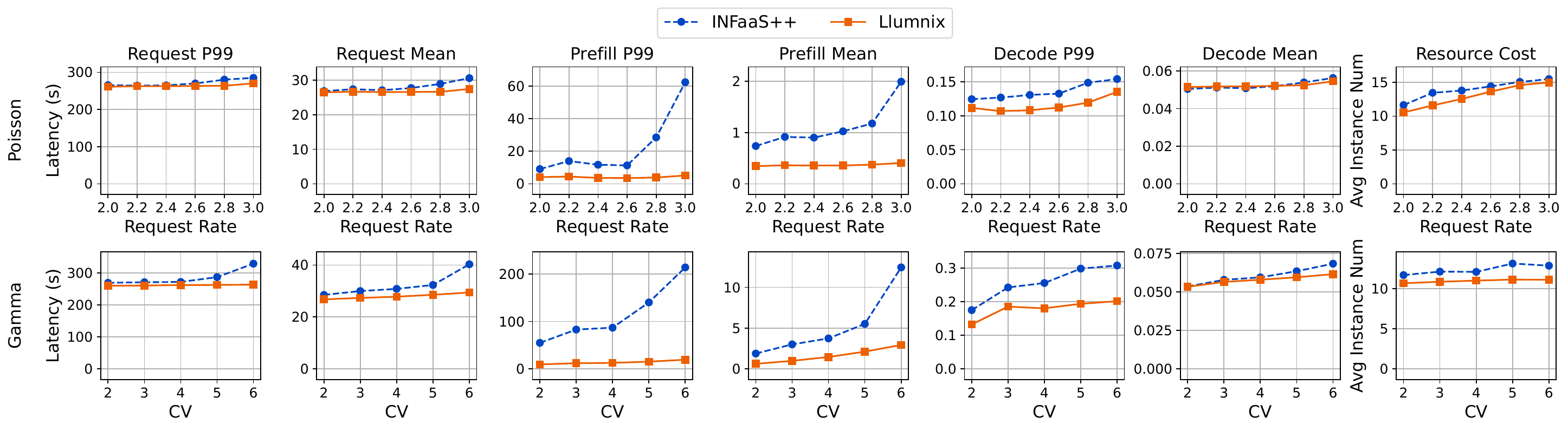}
    \caption{Auto-scaling of LLaMA-7B instances with Poisson and Gamma distributions, as annotated on the Y-axis labels.}
    \label{fig:eval-scaling}
\end{figure*}

\subsection{Auto-scaling}
\label{sec:auto-scaling}
We evaluate the auto-scaling capability of \sysname{} using larger ranges of request rates and Gamma CVs to show the adaptivity to load variation. By default, \sysname{} uses a scaling threshold range of [10, 60], \ie, \sysname{} scales instances up or down when the average freeness is under 10 or above 60; recall that this metric represents the most decode steps an instance can still run for given the current batch. We let INFaaS++ use the same scaling strategy, thus both \sysname{} and INFaaS++ have the same degree of aggressiveness of scaling up instances. We use a maximum instance number of 16 and the Long-Long sequence length distribution.

We first vary the request rates using Poisson distribution. As shown in the first row of Figure~\ref{fig:eval-scaling}, \sysname{} consistently achieves latency improvements across all request rates, \eg, up to 12.2$\times$ for P99 prefill latency. We also measure the resource cost in terms of average instances used, shown in the rightmost column.
\sysname{} saves costs by up to 16\%, because \sysname{} increases the auto-scaling efficiency by saturating or draining out instances more quickly. We also test different workload burstiness with varying CVs of Gamma distribution (request rate = 2). As shown in the second row, \sysname{} shows similar improvements in latencies and costs, \eg, up to 11$\times$ for P99 prefill latency and 18\% for the cost.

\begin{figure}[t]
    \centering
    \includegraphics[width=0.6\columnwidth]{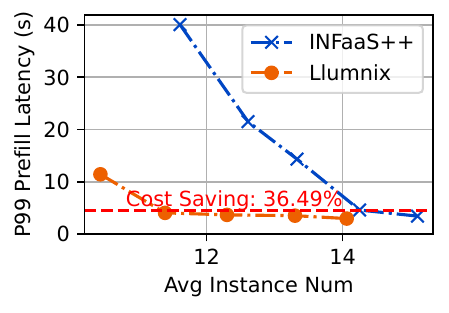}
    \caption{P99 prefill latencies vs. average numbers of instances with varying scaling thresholds.}
    \label{fig:eval-scaling-2}
\end{figure}

Finally, we examine the cost efficiency of \sysname{} in terms of how aggressively \sysname{} needs to scale out instances to preserve a certain latency objective, \eg, a given P99 prefill latency. We vary the scaling up threshold $t$, and the scaling threshold range is determined as [$t$, $t$+50]. Higher values of $t$ means that \sysname{} tends to use more instances. Figure~\ref{fig:eval-scaling-2} shows the P99 prefill latencies and costs with different scaling thresholds. We observe that \sysname{} achieves similar P99 prefill latency (roughly 5s, the red dash line) while saving 36\% of the cost compared to INFaaS++, as a result of the combination of the ability to reduce queuing delays via migration and the higher auto-scaling efficiency.

\subsection{Scheduling Scalability}

\final{
We conduct a scheduling stress test to examine the scalability of \sysname{} with 64 LLaMA-7B instances using higher request rates. Since this cluster exceeds the size of our testbed, we replace the real GPU execution in vLLM with a simple \texttt{sleep} command, whose duration is determined by offline measurement on A10 GPUs with different sequence lengths and batch sizes.
We build a simple centralized scheduler as the baseline by extending the vLLM scheduler to manage all requests across all instances. We issue requests with input and output lengths of 64 tokens with increasing request rates.
}

\begin{figure}[t]
    \centering
    \includegraphics[width=0.88\columnwidth]{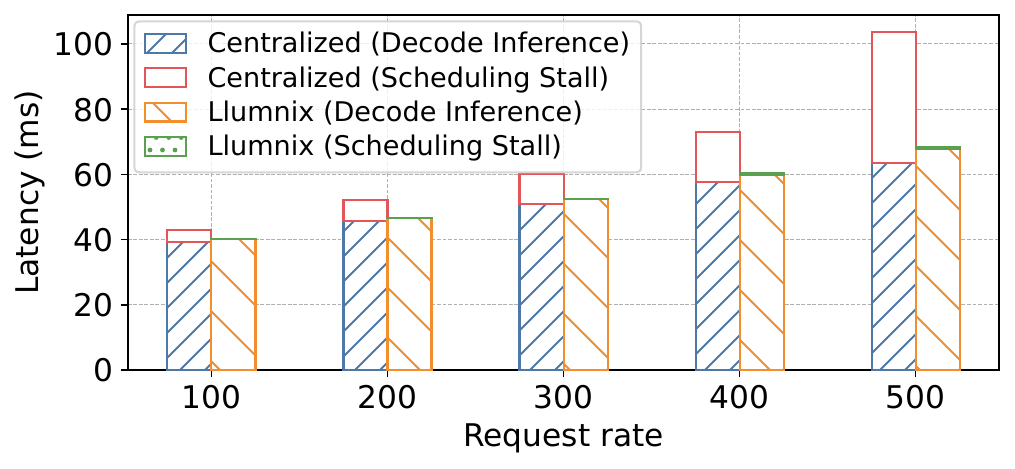}
    \caption{Per-token latencies and scheduling stalls under increasing request rates using 64 LLaMA-7B instances.}
    \label{fig:eval-scalability}
    \vspace{-10pt}
\end{figure}

\final{
As shown in Figure \ref{fig:eval-scalability}, with increasing request rates, the baseline experiences scheduling stalls during the inference computation of up to 40ms per iteration, translating to 1.7$\times$ slowdown. Such stalls are a result of the communication between instances and the centralized scheduler synchronizing request statuses and scheduling decisions, which becomes a bottleneck under high load. By contrast, \sysname{} exhibits near-zero scheduling stalls even under high request rates, showing the scalability of the distributed scheduling architecture. \sysname{} offloads and distributes the intra-instance scheduling logic across \syslet{}s so that it is done in parallel and asynchronously with the global scheduling. Moreover, \syslet{}s only report instance-level metrics, instead of the precise status of every single request, further improving the communication efficiency.
}

\section{Related Work}
\label{sec:related}
\paragraph{LLM inference.}
As transformer models show significance in model serving, recent works, such as FasterTransformer~\cite{faster2023},  TurboTransformer~\cite{turbotranformer2021}, LightSeq~\cite{lightseq2021}, and FlashAttention~\cite{dao2022flashattention, flashattention2}, optimize GPU kernels to improve the inference performance.
SpotServe~\cite{miao2023spotserve} supports LLM inference using preemptible instances for improving cost efficiency.
FastServe~\cite{wu2023fast} optimizes request completion times using a preemptive time-slicing approach.
AlpaServe~\cite{alphaserve2023} exploits pipeline parallelism to reduce serving latency for bursty workloads.
To further increase the GPU utilization and serving throughput, 
Orca~\cite{orca2022} proposes iteration-level scheduling (referred to as continuous batching in recent works and this paper) and selective batching, while vLLM~\cite{vllm2023} optimizes the memory usage with {PageAttention}. 
\cite{sheng2023fairness} proposes fair scheduling of requests on an LLM instance.
Prior works mostly target solo-instance serving, therefore complementing to \sysname{}. 
\sysname{} explores the challenges and opportunities of deploying multi-instance LLM serving. 
The key append-only characteristic of KV cache is exploited to enable migration capability of requests in the inference engine.
Such a mechanism opens great policy design space to offer priority and performance isolation, improve memory efficiency, and enable instance auto-scaling.
\revise{We also plan to explore the interplay between the global scheduling across instances with local scheduling techniques inside each instance (\eg, preemptive~\cite{wu2023fast} and fair~\cite{sheng2023fairness} scheduling) as future works.}












\paragraph{Request scheduling.}
To support deep learning model deployment, numerous systems (\eg, Clipper~\cite{clipper2017}, Nexus~\cite{nexus2019}, DVABatch~\cite{dvabatch2022}, and TritonServer~\cite{triton}) have been proposed to optimize request scheduling for DNN inference serving.
To meet the SLOs of DNN inference requests, 
Clockwork~\cite{clockwork2020} utilizes the execution predictability of traditional DNNs, while
Reef~\cite{reef2022} and Shepherd~\cite{shepherd2023} perform preemptions to serve high-priority requests. AlpaServe~\cite{alphaserve2023} uses a simple load-balancing dispatching policy based on queue lengths.
These works mostly focus on traditional DNN model serving, where a request requires only one-time inference on the model.
However, LLM inference service requires autoregressive computation on models for unpredictable numbers of iterations and introduces intermediate states (\ie, KV cache), showing brand new characteristics. DeepSpeed-MII~\cite{deepspeedmii}, albeit targeting multi-instance LLM serving, uses a simple round-robin dispatching policy that ignores LLM characteristics.
\sysname{} steps further to incorporate request migration and
ensures high throughput and low latency, provides SLO for prioritized requests, and auto-scales instances for resource efficiency with a unified load-aware dynamic scheduling policy.

\revise{Beyond multiple model instances, INFaaS~\cite{romero2021infaas} further supports scheduling across multiple model types/variants, considering the performance and accuracy requirements in difference applications. This is also a typical scenario for LLMs: for example, fine-tuned models for a specific task (\eg, coding~\cite{codellama,qwen,deepseek-coder}); variants with different sizes or precisions (\cite{frantar2023optq,ma2024era,lin2023awq}) of the base LLM. We plan to extend \sysname{} to support multiple model types in future work, considering the larger tradeoff space of latency/throughput and accuracy.}

\paragraph{Isolation vs. fragmentation.} 
\revise{
The tradeoff between isolation and fragmentation, or that between workload packing and spreading, have been a classic scheduling challenge. That is, workload packing improves resource utilization, at the expense of potential interference between co-located workloads; spreading workloads, on the contrary, provides better isolation but also increases resource fragmentation.
Many research efforts have been devoted to better balancing isolation and fragmentation in datacenters for big-data jobs and virtual machines, by identifying the interference-sensitivity of workloads and optimized scheduling policies (\cite{paragon,quasar2014,borg,mars2011bubble,yang2013bubble,hadary2020protean,cortez2017resource,han2016interference,chen2017prophet,garefalakis2018medea}).
This challenge was also identified in GPU clusters for deep learning workloads.
Amaral et al proposed a topology-aware placement algorithm to address the tradeoff between packing and spreading deep learning training jobs on multi-GPU servers~\cite{topoaware17}. Gandiva~\cite{gandiva2018} addresses the heterogeneous sensitivity to packing/spreading of different jobs with introspective job migration. This challenge becomes more complex for LLM serving due to the unpredictable autoregressive execution. \sysname{} exploits request migration at runtime to react to the workload dynamics to better reconcile these two goals.
}

\paragraph{Migration.}
Gandiva~\cite{gandiva2018} enables introspective migration for deep learning training jobs during scheduling. It utilizes the inherent iterative behavior of deep learning, and conducts checkpoint-resume approach on the minimal working set (\ie, mini-batch boundary) to migrate model weights. 
Even though LLM inference is iterative as well, directly migrating the entire states of a request is unacceptable, because the latency SLO of an inference request is crucial. Moreover, the working set per request is linear to the sequence length, which can be considerable given the trend of longer contexts~\cite{openai2023gpt4, gpt4turbo2023}.
The migration approach in \sysname{} is inspired by 
virtual machine live migration~\cite{vmmigration2005}.
By carrying out the majority of migration while LLM requests continue decoding tokens on GPUs, \sysname{} minimizes the downtime of request migration, making the cost negligible regardless of the sequence lengths.

\section{Conclusion}

\sysname{}, as implied by the name, represents our vision of serving LLMs as Unix. This vision originates in the observation that LLMs and modern operating systems have common natures such as the universality, multi-tenancy, and dynamism, and hence share similar requirements and challenges. This paper takes an important step towards this vision by drawing lessons from conventional OS wisdom including: definition of classic abstractions like isolation and priorities in the new context of LLM serving; implementation of the ``context switching'' as the key approach with inference request migration; and continuous, dynamic request rescheduling exploiting the migration. All these combined, \sysname{} delivers better latency, cost efficiency, and support for differentiated SLOs, pointing to a new way of LLM serving.

\section*{Acknowledgement}
We thank the anonymous OSDI reviewers and our shepherd for their valuable feedback, which helped improve the presentation of this paper. We thank Shiru Ren for early discussion on VM live migration techniques.
This work was supported by Alibaba Group through the Alibaba Research Intern Program. This work was also supported by National Key Research and Development Program of China (Grant No. 2023YFB3001801), National Natural Science Foundation of China (Grant No. 62322201, 62072018, U23B2020 and U22A2028), Fundamental Research Funds for the Central Universities (YWF-23-L-1121), and State Key Laboratory of Software Development Environment (SKLSDE-2023ZX-05).

\bibliographystyle{plain}
\bibliography{ref}

\newpage
\appendix
\section{Artifact Appendix}

\subsection*{Abstract}

This artifact includes the source code and scripts to run the experiments and reproduce the evaluation results of this paper.


\subsection*{Scope}


The artifact can be used to reproduce the results of the following experiments.

\begin{packed_itemize}
  \item Migration efficiency: Figure \ref{fig:microbench}.
  \item Serving performance: Figure \ref{fig:eval-serving}.
  \item Support for priorities: Figure \ref{fig:eval-priority}.
  \item Auto-scaling: Figure \ref{fig:eval-scaling} and Figure \ref{fig:eval-scaling-2}.
\end{packed_itemize}

\subsection*{Contents}


This artifact includes the following contents.

\begin{packed_itemize}
  \item Source code of a prototype implementation of \sysname{}.
  \item Scripts to prepare the environment, run the experiments, plot the figures, and validate the claims in this paper automatically.
  \item A \texttt{README} file including detailed instructions on how to use this artifact.
\end{packed_itemize}

\subsection*{Hosting}


The artifact is publicly available at \url{https://github.com/AlibabaPAI/llumnix} (the \texttt{osdi24ae} branch). Note that this is not the same branch as the official release of \sysname{} (the \texttt{main} branch). We will describe their difference later.

\subsection*{Requirements}


The artifact runs on GPU machines, with software dependencies mostly the same as those of vLLM.
To reproduce our results, you would need 4 GPU machines each with 4 A10 GPUs (24 GB). We recommend that you use the same VM type as in our experiments (\texttt{ecs.gn7i-c32g1.32xlarge} on Alibaba Cloud). 

\subsection*{Difference from the Official Release}


This artifact is a research prototype and was used during the experiments of this paper. After the paper submission, we refactored it into a new implementation that is more production-ready, \ie, the official release, as described in \S\ref{sec:impl}. Major differences between the two versions include:

\begin{packed_itemize}
  \item The artifact is directly based on the vLLM code base, whereas the official release is a standalone Python library, making it more extensible and non-intrusive to backend inference engines.
  \item The artifact is not fault-tolerant, whereas the official release provides fault tolerance for each component.
  \item \final{The official release is still being actively developed, and has supported or will support a series of new features, such as scalable API servicing via distributed request frontends, support for newer versions of vLLM and more models, further improvements of the scheduling policies, \etc.}
\end{packed_itemize}

The artifact is sufficient to reproduce the experiment results in this paper. However, if you want to use \sysname{} in production or conduct further research, we do recommend the official release.


\end{document}